\documentclass[aps,prb,twocolumn,groupedaddress,showpacs,letterpaper]{revtex4}

\usepackage[dvips]{graphics}

\begin{document}
\title{Dynamical signature of a domain phase transition
in a perpendicularly-magnetized ultrathin film}
\author{N. Abu-Libdeh and D. Venus} 
\email{venus@physics.mcmaster.ca}
\affiliation{Dept. of Physics and Astronomy, McMaster University,
Hamilton Ontario, Canada} 
\date{\today}

\begin{abstract}
Domain phases in ultrathin Fe/Ni/W(110) films with perpendicular
anisotropy have been studied using the ac magnetic susceptibility.
Dynamics on time scales of minutes to hours were probed by quenching
the system from high temperature to the stripe phase region, and
varying the constant rate of temperature increase as the
susceptibility traces were measured. The entire susceptibility peak is
observed to relax slowly along the temperature axis, with the peak
temperature increasing as the rate of heating is decreased.  This is
precisely opposite to what would happen if this slow relaxation was
driven by changes in the domain density within the stripe phase.  The
data are instead consistent with a simple model for the removal of a
significant density of pattern defects and curvature trapped in the
quench from high temperature.  A quantitative analysis confirms that
the relaxation dynamics are consistent with the mesoscopic
rearrangement of domains required to remove pattern defects, and that
the experiment constitutes a ``dynamical'' observation of the phase
transition from a high temperature, positionally disordered domain
phase to the low temperature, ordered stripe phase.
\end{abstract}
\pacs{75.70.-i, 75.30.Kz, 75.40.Cx}
\maketitle

\section{Introduction} 
The study of domain pattern formation in two-dimensional systems with
strong short-range attractive interactions and weak, long-range
repulsive dipole interactions links disparate fields such as molecular
membranes,\cite{sagui} crystals exhibiting high temperature
superconductivity,\cite{kivelson} and ultrathin film
magnetism.\cite{debell} In all these cases, the phase diagram is
influenced by the strong fluctuations in two dimensions, which lead to
novel phases and phase transitions.  The domain patterns of ultrathin
ferromagnetic films with weak perpendicular anisotropy have been
studied intensely because this system is amenable to precise
experimental control.  Even so, the predicted phases and phase
transitions have proven difficult to observe because imaging
techniques average over the fluctuations.  The purpose of the present
article is to report observations of the slow relaxation of domain
patterns that have been quenched from high temperature, using the
ac magnetic susceptibility.  The relaxation is consistent with the
resolution of pattern defects as the low temperature stripe phase is
formed.

Ultrathin films possess a surface magnetic anisotropy that may favour the 
alignment of the magnetic moments normal to the surface.  In this case, a 
demagnetization field can cancel much of the surface anisotropy, leaving a 
weak, temperature-dependent, residual perpendicular anisotropy.  The 
resulting small domain wall energy permits magnetic domain patterns to 
form despite the weakness of the long-range dipole interaction.  Numerous 
theoretical analyses\cite{yafet,kashuba,abanov,politi} and computer 
simulations\cite{booth, macisaac, stoycheva,sampaio} indicate that a 
``stripe'' pattern is stable at low temperature.  Great progress has been 
made in understanding this phase using magnetic microscopy 
techniques.\cite{allenspach,speckmann,won,portmann,saratz} At higher 
temperature (lower anisotropy), fluctuations in the domain walls become 
important, and stripe domains may meander and ``pinch off" to create pairs 
of dislocations.  As the dislocations proliferate, theory 
predicts\cite{kashuba,abanov} a Kosterlitz-Thouless transition to a domain 
phase that is characterized by pattern defects, domain segmentation, and 
the loss of long-range positional order.  At even higher temperature, the 
loss of orientational order, and finally paramagnetism are predicted.  As 
a group, these phases may be termed ``positionally disordered'' phases.  
Only a few experiments have produced magnetic images showing relevent 
defect structures.\cite{vaterlaus,portmann2} Caution is required in 
interpreting these essentially static images of phases driven by 
fluctuations.  For example, images showing a loss of domain contrast may 
indicate a transition to the paramagnetic state, or simply a dynamic 
effect that averages out the image contrast.

A complementary method to study the domain phases is through their
dynamics.  This offers access to a wide range of time scales where
different relaxation processes are active, at the expense of a
pictorial representation of the domain pattern.  There are very few
studies of the intrinsic dynamics of these magnetic systems.  Early
work concentrated on the relaxation from the magnetically saturated
state.\cite{berger} Work measuring the magnetic susceptibility studied
the transition from the stripe phase to a ``glassy'' stripe phase due
to pinning of the domain walls by structural
defects.\cite{venus,venus2} More recently, numerical simulations have
suggested that the transitions between the various phases have
distinctive dynamic signatures. Starting from a magnetically saturated
initial state, simulations find very long equilibration times for the
creation of the stripe phase, but much shorter equilibration times for
the positionally disordered phase.\cite{bromley} Simulations of
quenching from a positionally disordered state indicate a
discontinuous transition with long-lasting metastability before the
stripe phase is established.\cite{cannas}

The present article reports experimental studies of ultrathin
Fe/Ni/W(110) films aimed at quantifying the dynamics of domain pattern
relaxation across a range of time scales by altering the constant rate
of temperature change, R, as the magnetic susceptibility is scanned.
Following a controlled quenching from high temperature, the evolution
of the susceptibility shows that the system as a whole slowly relaxes
towards equilibrium through activated dynamics.  It is then argued
that this does not represent the relaxation of the stripe domain
density (which occurs on a shorter time scale), but is rather due to
the activation barriers for the mesoscopic domain wall rearrangements
required for the removal of dislocations and defects as the quenched
pattern moves toward a low temperature equilibrium stripe phase.  This
represent a ``dynamical'' observation of the transition from a
positionally disordered phase to the equilibrium, stripe domain phase.

\section{Theory}
The competition between the short range exchange interaction and the
long range dipole interaction in a two dimensional magnetic system
with perpendicular anisotropy leads to the spontaneous formation of
magnetic domains.  A detailed study of a two dimensional system of
domain walls, in the presence of a substrate that induces preferred
directions of domain wall alignment, has outlined the expected phase
diagram of the domain phases.\cite{kashuba,abanov,politi} At low
temperature, the system forms positionally and orientationally ordered
stripes (also termed the ``smectic'' phase) with a mean linear density
$n_{eq}(T)$
\begin{equation}
n_{eq}(T)=\frac{4}{\pi \ell} \exp(-\frac{E_W}{4\Omega N}-1),
\end{equation}
where $\ell$ is the domain wall width, $E_W$ is the domain wall energy
per unit length, $\Omega$ is a constant that sets the scale of the
dipole energy, and $N$ is the number of monolayers in the thin film.
The residual perpendicular anisotropy is reduced by spin fluctuations,
so that the domain wall energy depends sensistively on temperature.
This causes the domain density to increase exponentially with
temperature.

When a magnetic field is applied perpendicular to the surface, the
domains with moments parallel to the field grow at the expense of
those with moments in the antiparallel direction, creating a net
moment.  In the stripe phase, where the bending of the domain walls is
unfavourable, the equilibrium dc susceptibility $\chi_{eq}(T)$ is
determined by the induced changes in domain wall spacing, and is
directly related to the effective spring constant for stripe
compression,\cite{abanov,venus} $k_{eff} \sim 1/\chi_{eq}$, where
\begin{equation}
\label{suscept}
\chi_{eq}(T)=\frac{4}{\pi d n_{eq}(T)} \sim A\exp (-\kappa T).
\end{equation}
$d$ is the film thickness.  The phenomenological parameters $\ln
A$ and $\kappa$ represent the zeroth and first order expansion of the
exponential dependence of the domain density on temperature.  

When an ac field is applied, the oscillating motion of the domain
walls occurs through Barkhausen steps of microscopic lengths of domain
wall between thermally activated pinning sites with time
constant\cite{bruno}
\begin{equation}
\label{tau}
\tau_{pin} (T) = \tau_{0pin}\,  \exp (T_{pin}/T),
\end{equation}
where $T_{pin}$ is the pinning energy.  Solution of a relaxation
equation for the magnetization measured at angular frequency $\omega$
gives the ac susceptibility as
\begin{equation}
\label{chieff}
\chi(T)=\frac{1-i\omega\tau_{pin}(T)}{1+\omega^2\tau_{pin}^2(T)} \,
\chi_{eq}(T).
\end{equation}
The susceptibility falls exponentially with temperature on either side
of a maximum.  At low temperature the domain wall motion is stopped by
pinning, and at high temperature by the increasing magnetic stiffness
of the domain walls as their density increases.  This characteristic
shape has been observed in many studies, and permits a quantitative
study of the pinning mechanism.\cite{venus,venus2,arnold}

As the temperature is increased, fluctuations in the domain walls are
favoured, and dislocations pairs are formed by segmentation of the
stripe domains.  According to the continuum model of Abanov \emph{et
al.}\cite{abanov}, these dislocations proliferate at a
Kosterlitz-Thouless transition to the ``Ising nematic'' domain phase
that is characterized by unbound dislocations that destroy the long
range positional order.  This (or a related positionally \emph{and}
rotationally disordered phase) exists for a small temperature range as
the exponentially decreasing domain width approaches and reaches a
fundamental limit, given by the dipole length, or ratio of exchange
and dipole energies.\cite{kashuba} Near this fundamental limit, the
shape of the domain wall is altered\cite{vedmedenko,vindigni} and the
width becomes equivalent to the domain width, so that the continuum model
is no longer strictly valid.  At even higher temperature, there is a
transition to in-plane magnetism, or paramagnetism, depending upon the
sample thickness.

Numerical simulations of a 2 dimensional layer of Ising spins are in
agreement with many qualitative features of this
description\cite{booth, macisaac, stoycheva,sampaio}, particularly in
observing stable domain patterns with the properties of the ordered
stripe phase and of disordered phases containing defects.  More
recently, simulations have been used to study the relaxation dynamics
and domain phase transitions in Ising systems.  For relaxation from a
magnetically saturated state, to the stripe phase, Bromley \emph{et
al.}\cite{bromley} find three distinct relaxation rates: i) for the
formation of domain segments, ii) for the establishment of a
symmetrical distribution of up and down domains giving no net
magnetization, iii) for the alignment of the domains along a single
stripe direction.  Particularly the last step leads to a very long
equilibration time.  In contrast, the same study finds a much faster,
single step relaxation to the positionally and rotationally disordered
phase.  In simulations of quenching the positionally disordered phase
into the stripe phase region, Cannas \emph{et al.}\cite{cannas} found
more complex transitions than are predicted by the continuum model.
In particular, their simulations indicate a discontinuous transition
between the stripe and nematic phases, rather than a
Kosterlitz-Thouless transition.  The discontinuous transition is
accompanied by temperature hysteresis, and long lasting metastability
of the disordered phase after quenching.

Both of these numerical studies of Ising systems suggest that these
distinctive dynamic responses may permit an indirect observation of
different domain phases in experimental systems with realistic domain
wall widths.  Measurements of the magnetic susceptibility should be
sensitive to relaxation dynamics on a number of time scales, but it is
not certain how the magnetic susceptibility of the stripe phase would
differ from a system with substantial domain wall curvature and
defects, such as the disordered domain phases.  However, since the
disordered phases retain short range order, and because the domain
density is expected to increase with temperature irrespective of the
phase, the ac susceptibility should continue to be described, at least
\emph{qualitatively}, by eq.(\ref{chieff}).  On the other hand, when
domain wall fluctuations, curvature and dislocations become important
and contribute to the elastic energy of the system, one should expect
\emph{quantitative} changes in the effective spring constant
$k_{eff}$, and therefore in the susceptibility.

\section{Experimental results}
Films of 1.5 ML Fe/2.0 ML Ni/W(110) were grown in ultrahigh vacuum and
studied \emph{in situ}.  The growth procedures were taken from
previous studies of their structural and magnetic
properties.\cite{johnston} The films have perpendicular anisotropy at
low temperature.  For films thicker than 2.2 ML, there is a
re-orientation transition from perpendicular to in-plane moments as
the temperature is increased, followed by a Curie transition to
paramagnetism.\cite{arnold} The film thickness, cleanliness and
structure were confirmed using Auger electron spectroscopy and low
energy electron diffraction.  ac magnetic susceptibility measurements
at a frequency of 210 Hz using a field amplitude of 2.0 Oe were made
using the surface magneto-optical Kerr effect, and lock-in
amplification.\cite{arnold2} The sample temperature was measured using
a W-Re5\%/W-Re10\% thermocouple, and was controlled by radiant heating
from a filament, and conductive cooling through a copper braid
attached to a liquid nitrogen reservoir.  When heating, the average
rate of change of the sample temperature, R, could be controlled from
0.05 K/s to 1.00 K/s, with fluctuations in the rate less than 0.05 K/s.
The maximum rate of controlled cooling that could be maintained over
the entire relevant temperature range was -0.10 K/s.

A number of steps were taken to ensure reproducible measurements.
First, the films were annealed to 400K after growth, and subsequent
measurements did not exceed 360 K.  In each case where the
susceptibility was measured by heating from low temperature, the
sample was first cooled from 360 K at a rate of R = -0.10 K/s.  Since
the initial cooling of the sample and sample support was much slower,
the first measurement upon heating was discarded.  When a series of
curves were measured from the same film at different heating rates,
the order of the rates was randomized so that effects of aging could
not masquarade as effects due to changes in the heating rates.
Finally, the effect of changes in the time constant used for lock-in
amplification in the experiments was studied.  In this way the time
constant was chosen as 2 s.  This is as large as possible to reduce
noise, yet smaller than the value that affected the shape of the
experimental curve.

\begin{figure} 
\scalebox{.4}{\includegraphics{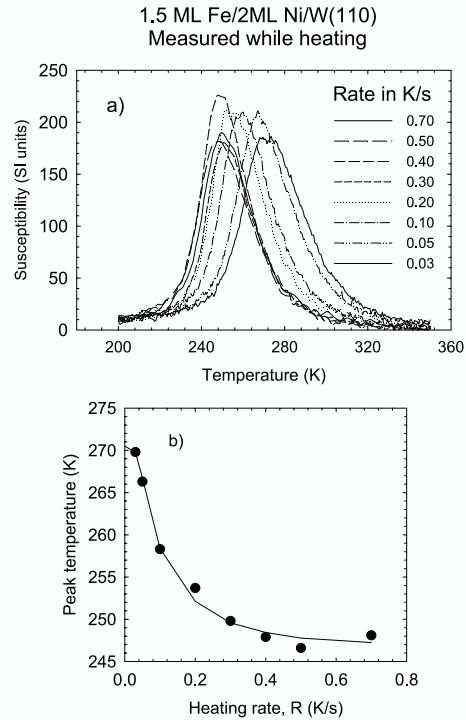}} 
\caption{Magnetic susceptibility of a 1.5 ML Fe/2.0 ML Ni/W(110) film
measured at different constant rates of temperature change, R. a)
Measurements for heating the film, taken after cooling from 360 K at
-0.10 K/s.  b) The temperature where the susceptibility peaks, as a
function of the heating rate.  The fitted line is discussed in section
IV.}
\end{figure}

The real part of the ac magnetic susceptibility measured at a series
of heating rates on a single film is shown in fig. 1a.  The shape of
the curve changes only in subtle ways as the heating rate is changed
-- the primary effect of changing the heating rate is a shift of the
entire susceptibility curve in temperature.  In the range 0.70 K/s
$\geq R \geq $ 0.30 K/s, there is very little shift, but for slower
heating rates the peak shifts progressively to higher temperature, and
the susceptibility curve broadens somewhat.  The location of the peak
as a function of the heating rate is summarized in fig. 1b.

\begin{figure} 
\scalebox{.4}{\includegraphics{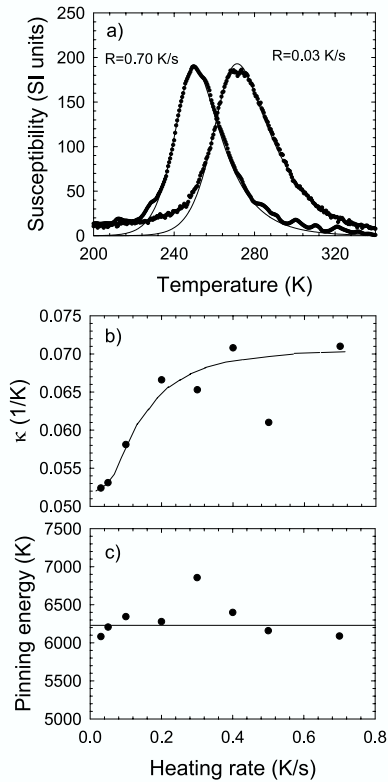}} 
\caption{a) Representative fits of the susceptibility curves in
fig. 1a to eq.(\ref{chieff}). b) The parameter $\kappa$ in the fit.
The line is a guide to the eye.  c) The parameter $T_{pin}$ in the
fit.  The line is the average value.}
\end{figure}

The changes in the susceptibility curve upon heating have been
characterized by fitting the curves to
eq.(\ref{suscept}-\ref{chieff}).  The fits are summarized in fig. 2.
In part a), the solid lines show the fits to the data for the extreme
values of $R$ from fig. 1.  These fits are representative, with the
curve following the data closely except at the low temperature tail.
The low temperature tail deviates from the data because the fit
assumes a single average pinning energy, $T_{pin}$, rather than a more
correct distribution of pinning energies.  There are relatively few
pinning sites with a pinning energy significantly lower than the
average, but these are the only sites where the domain walls remain
free at low temperature, and thus account for all the signal in the
tail of the curve.\cite{venus2} The greater size of the low frequency
noise in the data for $R=0.70$ K/s is consistent with the reduced
measurement time compared to the data with $R=0.03$ K/s. Fig. 2b and
2c present the fitted parameters $\kappa$ and the average defect pinning
energy $T_{pin}$, respectively.  It is clear that the shift in the
peak temperature is associated with a change in $\kappa$.  The
pinning energy remains almost constant, as is expected, since
the pinning energy is a characteristic of the film and substrate
microstructure and does not change with the heating rate.

Since $\kappa$ is an effective parameter, it is difficult to be
precise about what causes it to change with heating rate so that the
peak shifts in temperature.  However, a number of possible artifacts
can be ruled out.  Fundamentally, changes in the equilibrium domain
patterns (and thus the susceptibility) are driven by changes in the
perpendicular anisotropy as a function of temperature. However, the
perpendicular anistropy is determined by materials properties
(magneto-crystalline anisotropy, shape anisotropy) that are not
dependent on the \emph{rate} of temperature change.  The peak shifts
are therefore not due to a true shift in the temperature dependence of
the perpendicular anisotropy.  It is, however, conceivable that the
peak shifts are due to a time lag between the measured and true
temperature, or due to the response time of the temperature
measurement system itself.  A simple calculation of the thermal
diffusion time across the tungsten substrate gives an upper limit of a
few tenths of K for a shift due to the fact the temperature
measurement point and experimentally probed area of the film are
physically separated.  Limits on the size of the thermal lag of the
overall temperature measurement system were established by heating at
a constant rate up to a specific temperature, and then monitoring the
measured temperature as a function of time at a constant temperature.
In this case, a conservative upper limit in the temperature shift of 1
or 2 K was established.  These effects are much smaller than those
shown in fig. 1b.

It can also be shown that the changes in $\kappa$ with heating rate
are not due to the dynamical processes involved in changes to the
domain density with temperature.  At low temperature, the domain
density will be small.  As the temperature is increased, the domain
density will increase through the creation and growth of new stripes.
Since this process is thermally activated, the domain density will
always lag the equilibrium domain density.  As the heating rate is
increased, this effect will be more pronounced, and the amount of lag
between the actual and equilibrium density at a given temperature will
also increase, and this could affect the fitted value of $\kappa$ and
shift the susceptibility curve.  However, the crucial point is that,
during heating, the lagging domain density is smaller than the
equilibrium value, so the measured susceptibility is increased,
causing the measured susceptibility curve to be progressively shifted
to higher temperature for higher heating rates.  This is precisely the
\emph{opposite} of what is seen experimentally.  Although changes in
domain density must occur, apparently this dynamical process is active
on a faster time scale than those probed by the present experiments on
these samples.  On the time scale of the present measurements, the
system is able to continually maintain a domain density close to the
equilibrium value, except perhaps at the lowest temperatures.  This
conclusion is further corroborated by the very good fits of the
susceptibility curves to eq.(\ref{chieff}) reflecting the underlying
processes of domain wall pinning at low temperature and a rapid
increase in domain density at high temperature.  The shift in the peak
temperature in fig. 1 must be predominantly due the relaxation of some
quantity other than domain density.

\begin{figure}
\hspace{-.7in}
\scalebox{.4}{\includegraphics{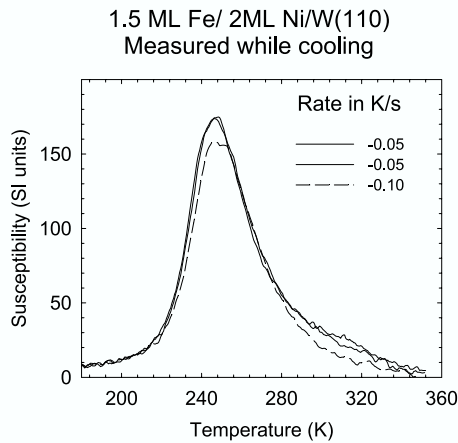}}
\caption{The magnetic susceptibility measured while the sample was
cooled at a few different rates.}
\end{figure}

A series of further experiments were performed to more completely
characterize the system.  The susceptibility measured for a different
sample during cooling is presented in fig. 3.  Even though the range
of $R$ available for cooling is limited, it is clear that the
difference in the curves for $R$ = -0.10 K/s and -0.05 K/s is at most
small and likely negligible.  There is a distinct asymmetry in the
behaviour for heating and cooling at these rates.  The peak position
and shape for the cooling curves is most similar to those of the
heating curves at $R \geq $ 0.40 K/s.

\begin{figure}
\scalebox{.4}{\includegraphics{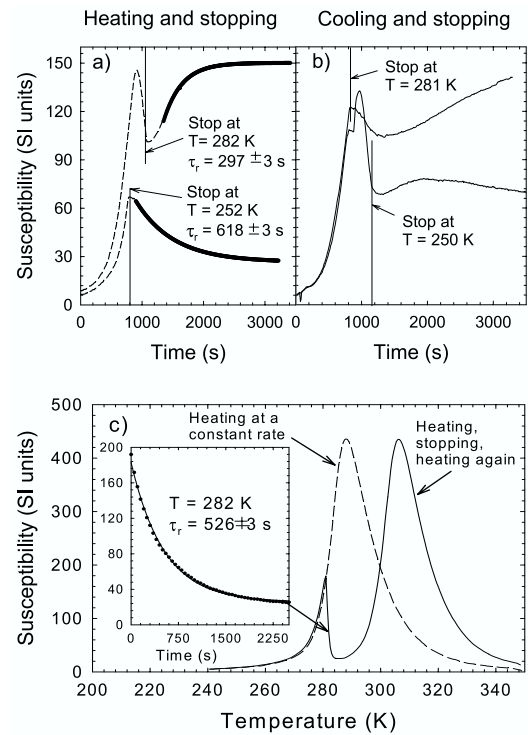}}
\caption{a) The magnetic susceptibility as a function of time,
measured while heating at 0.10 K/s. The heating was stopped at the
specified temperature, and then monitored as a function of time at
constant temperature.  The solid line is an exponential fit to the
relaxation.  b) As in a), except that the susceptibility is measured
while cooling at -0.10 K/s.  c) A different sample is heated at 0.30
K/s (dashed line), then cooled at -0.10 K/s and measured during
heating again (solid line).  This time the heating is stopped at 282
K, and the susceptibility is monitored as a function of time at
constant temperature (inset).  The heating is then recommenced,
producing the remainder of the curve.}
\end{figure}

Longer time scales have been probed by stopping the heating or cooling
at a predetermined temperature, and monitoring the susceptibility as a
function of time.  These results are presented in fig. 4a and 4b, for
heating and cooling respectively, with $|R|$ = 0.10 K/s.  After
heating to 252 K, the susceptibility relaxes to a lower value by a
simple exponential decay with $\tau_r=618\pm 3 $s, whereas after
heating to 282 K, the susceptibility relaxes more quickly
($\tau_r=297\pm3$s) to a higher value.  These results are in agreement
with those presented in fig. 1a; the relaxation is always toward the
curve measured with a smaller heating rate, which is closer to
equilibrium.  The curves in fig. 4b confirm the asymmetry between
heating and cooling.  When cooling is stopped at these same
temperatures, the long-term relaxation is in the same direction as for
heating (after a transient), but the relaxation time constant is much
larger.  Rough estimates are $\tau_r \approx 5,000$ s at 250 K and
$\tau_r \approx 1,200$ s at 281 K.

A final experiment on a third film is presented in fig. 4c.  In this
case, the susceptibility was measured for a heating rate of $R$ = 0.30
K/s (dashed curve), the sample was cooled once more and the
measurement was repeated (solid curve).  This time the heating was
stopped at $T$=282 K, and the relaxation of the susceptibility was
monitored as a function of time, as indicated in the inset.  After the
relaxation was complete, heating at $R$ = 0.30 K/s was resumed,
producing the remainder of the solid curve.  This result confirms that
the shift in the curves is not caused by the different heating rates
\emph{per se}, but rather that the system is relaxing from a state
that produces a susceptibility peak at lower temperature to one with a
peak at a higher temperature, and that the amount of relaxation depends 
upon the total time that has elapsed.

\section{Analysis and Discussion}

These results provide evidence that some quantity in the quenched
samples is relaxing as the equilibrium stripe phase is established,
and that this relaxation occurs on a longer time scale than that
required for the relaxation of the domain density.  This is
qualitatively consistent with the numerical results of Cannas \emph{et
al.}\cite{cannas} who suggest the high temperature, positionally
disordered smectic phase will be strongly metastable and persist into
the stripe phase.  It is also qualitatively consistent with Bromley
\emph{et al.}\cite{bromley}, who find that the resolution of an
initial state with many pattern defects and dislocations into an
ordered stripe pattern proceeds with a very long time constant. We
therefore advance the hypothesis that quenching the system traps a
configuration with significant domain curvature and density of domain
dislocations at low temperature, and that the system relaxes very
slowly to an ordered stripe pattern.  Upon heating the sample, the
experiments indicate a more efficient resolution of the pattern
defects during domain creation than was the case during domain
annihiliation (cooling).  This more efficient relaxation is revealed
by a shift of the susceptibility peak to higher temperature, and a
change in the phenomenological parameter $\kappa$, that is related to
domain formation.  It allows the system to approach the equilibrium
state at high temperature, so that the system shows reproducible and
systematic hysteresis upon temperature cycling with different heating
rates.

To test this interpretation, it is first assumed that the relaxation
rate during cooling is so much slower than during heating that the
amount of relaxation that occurs during cooling can be neglected.  The
relaxation during heating is modeled as an activated process with time
constant
\begin{equation}
\label{taur} 
\tau_r=\tau_{0r}\, \exp (T_r/T), 
\end{equation}
where $T_r$ is an activation barrier that must be overcome in relaxing
to the equilibrium stripe phase.  The total number of time constants
that have passed while heating from the initial temperature $T_i$ up
to the peak temperature $T_{peak}$ can be defined as $t_{eff}$:
\begin{equation}
\label{teff}
t_{eff}(R)=\int_{T_i}^{T_{peak}(R)} \frac{dT}{R\tau_r(T)}.
\end{equation}
Using the peak of the susceptibility curve as a marker to follow the
relaxation suggests
\begin{equation}
\label{relax}
T_{peak}(R)=T_0 -\Delta \exp (-t_{eff}(R)).
\end{equation}

\begin{figure}
\scalebox{.5}{\includegraphics{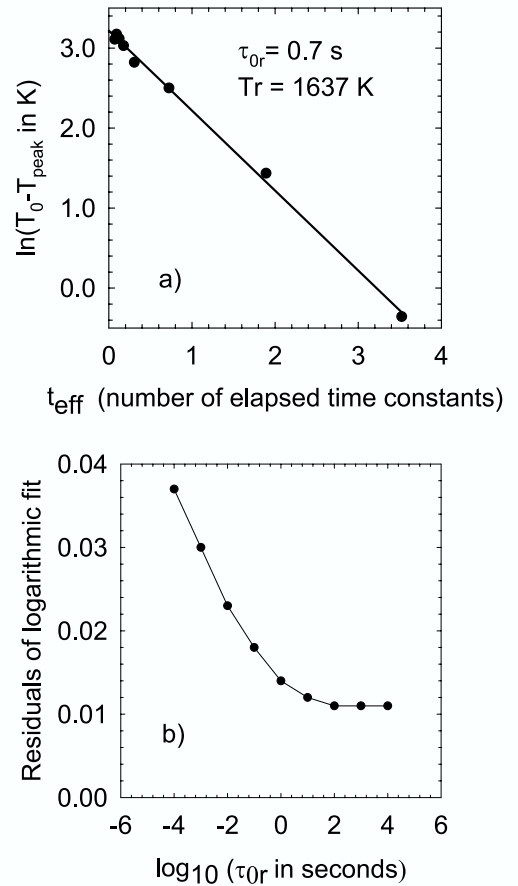}}
\caption{Fit of the relaxation of the temperature of the
susceptibility peak to eq.(\ref{relax}).  a) Peak temperatures plotted
agains $t_{eff}$ as defined in eq.(\ref{teff}), assuming
$\tau_{0R}=0.7$ s.  $T_R=1637$ K is the fitted value of the nucleation
energy for the phase change. $T_0=270.5$ K and $\Delta=25.0$ K. b) The
residuals of the logarithmic fit in part a), as a function of
$\log_{10} (\tau_{0r})$.}
\end{figure}

$T_0$ is the peak temperature when relaxation is complete.  A plot of
$\ln (T_0 - T_{peak})$ vs. $t_{eff}$ has a slope of -1, and intercept
of $\ln \Delta$.  There are three adjustable parameters, $\tau_{0r}$,
$T_r$, and $T_0$, but two are linked by the requirement that the slope
of the plot is -1, leaving two independent parameters.  The least
squares fit of the peak positions to eq.(\ref{relax}) is shown in
fig. 5a, and as the solid line in fig. 1b.  The value chosen for
$\tau_{0r}$ is constrained by the least squares residual in
logarithmic space, as shown in fig. 5b, which gives a lower limit of
$\tau_{0r} \approx 10^{0}$, but does not establish an upper limit.
However, the two experimental relaxation times in fig. 4a allow an
independent estimate of the two quantities that determine the time
constant in eq.(\ref{taur}) as $\tau_{0r}$ = 0.7 s, and $T_r$ = 1735 K.
This value of $\tau_{0r}$ is entirely consistent with the lower limit
established by the fitting, and was used for the plots in fig. 5a and
1b. The fitted value of $T_r = 1637 K$, corresponding to $\tau_{0r}$ =
0.7 s, is in very reasonable agreement with the experimental estimate
of 1735 K. It is clear that this simple analysis gives a very good
account of the data.

The excellent fit to the simple relaxation model for the peak as a
whole has a number of implications.  First, it confirms that
``quenching'' is an appropriate description of the cooling of the
system, since essentially no relaxation occurs during cooling.  The
measurement of $\chi(T)$ for $R = 0.70$ K/s in fig. 2a is, in some
important way, unrelaxed from the initial high temperature state
before quenching.  With $t_{eff}$ = 0.07 time constants, approximately
only 7\% of the relaxation has occured.  By contrast, the measurement
with $R = 0.03$ K/s has $t_{eff}$ = 3.52 time constants, so that 97\%
of the relaxation is complete.  The data at intermediate values of $R$
in fig. 1a are not a linear combination of these two endpoints -- if
they were, $\chi(T)$ would exhibit a double-peaked structure.  It
therefore seems that the relaxation is not related to a localized
property such as, for instance, phase separation or nucleation and
growth of a low-temperature phase.  Rather, a distributed property,
which is ultimately related to $\kappa$, is relaxing.  The analysis is
therefore consistent with the relaxation of a non-equilibrium density
of domain dislocations, and/or domain wall curvature that are
characteristic of high temperature state from which the quench
occurred.

The fundamental time scale of the relaxation, $\tau_{0r}$, is of order
1 s, which is very long compared to the fundamental time scale for
localized Barkhausen steps in the domain wall position, $\tau_{0pin}
\approx 10^{-9}$ s, in eq.(\ref{tau}).  This implies that the
relaxation requires changes of the domain walls on a mesoscopic scale
involving the co-ordinated action of many independent microscopic
Barkhausen steps, and is not directly related to the shape profile of
the domain wall.\cite{vedmedenko,vindigni} The activation energy, $T_r
\approx$ 1600 K, for this co-ordinated motion of many Barkhausen
steps, is about 1/4 of the activation energy $T_{pin} \approx 6200$ K
for the pinning of just one local Barkhausen step.  Taken together,
these indicate that the relaxation involves larger scale domain
rearrangments driven by an interaction that is intrinsically weak but
becomes substantial when integrated over a large area.  This again is
consistent with the relaxation of domain curvature and dislocations,
driven by dipole energies.

In addition, there is clearly a relation between the phenomenological
parameter $\kappa$ and the relaxation.  The larger values of $\kappa$
for susceptibility measurements with larger values of $R$ indicate
that after quenching, but before relaxation, the domain pattern is
magnetically stiffer -- that is, the effective spring constant for the
compression of the pattern by a magnetic field is significantly
larger.  In the equilibrium stripe phase at low temperature, the
magnetic stiffness depends only upon the domain density.  Curvature of
the domain walls and dislocations are essentially absent in
equilibrium because they would greatly increase the elastic energy.
At high temperature and in the positionally disordered phases, on the
other hand, the enhanced fluctuations of the domain walls can
accommodate curvature and an equilibrium density of domain
dislocations without a significant cost in elastic energy.  If
quenching from high temperature introduces a relatively large amount
of domain curvature and density of dislocations to the low temperature
domain pattern, it is reasonable that this would be reflected in a
larger elastic constant and value of $\kappa$, which would decrease,
as is observed in fig. 2b, as the relaxation to equilibrium removes
the defects.

Finally, there is the question of the asymmetry of the relaxation
times for heating and cooling the system. Since domain annihilation
upon cooling proceeds by first breaking up existing domains, it
creates additional pattern defects and works against the removal of
existing defects, making the relaxation very slow.  Domain creation
during heating, on the other hand, provides an opportunity for defects
to be removed by growth and merging with other defects, or at the very
least dilutes them in a greater stripe density.  This abets the
relaxation of the system toward equilibrium.

\section{Conclusions}

Significant hysteresis and a heating/cooling asymmetry has been
observed in the magnetic susceptibility of perpendicularly magnetized
1.5 ML Fe/2 ML Ni/W(110) films.  The entire susceptibility curve relaxes
slowly to higher temperature as the rate of heating during the
measurement is decreased.  This is incompatible with relaxation of the
domain density, which occurs on a faster time scale and causes a
temperature shift of opposite sign.  However there is a great deal of
self-consistent evidence that it is compatible with the relaxation of
dislocations, defects and curvature that have been quenched from high
temperature into an unstable domain configuration at low temperature.
This evidence includes; an excellent quantitative fit of the
relaxation to an activated relaxation with a fundamental time and
energy scales appropriate for a large scale domain rearrangement by
the weak dipole interaction; the correlation of the relaxation with a
reduction in the magnetic ``stiffness" of the film; and a qualitative
link between the annihilation/creation of domains and the
creation/annihilation of defects to understand the asymmetry in
heating and cooling .  This suggests that in quenching from 360 K to
220 K, the system passes through a transition from a positionally
disordered domain phase to the semectic stripe phase.

These observations give qualitative support to recent numerical
studies that indicate that the removal of domain dislocations after
moving across a phase transition to the ordered stripe phase is very
slow.  However, it is not possible to discriminate between the
detailed pictures suggested in those studies; whether the transition
is of the Kosterlitz-Thouless type or is discontinuous\cite{cannas},
whether it originates in the Ising nematic or tetragonal domain
phases\cite{bromley}, and whether the quenched system is metastable or
merely dynamically hindered.  Regardless of these details, the results
have important implications for the interpretation of studies using
imaging techniques to study these domain phases.  They suggest that
great care must be taken to obtain images of phases that represent
equilibrium conditions.  Most imaging studies do not give details of
the thermal history of the samples, although some mention very long
times for the evolution of some patterns.\cite{portmann} It is often
not clear that images that seem to show paramagnetic or tetragonal
phases are close to equilibrium and can serve as a basis for creating
a phase diagram.  It may be that the imaging of patterns with many
defects is possible only because of the very slow dynamics observed in
the present study.

The present dynamical studies highlight two different processes that
occur on very different time scales in perpdendicularly magnetized
domain patterns -- the motion of domain walls in response to an
applied field and the removal of pattern defects in response to long
range dipole interactions.  We expect that there will also be a
dynamical signature of the annihilation and creation of domains to
adjust the domain density on an intermediate time scale, and are
currently conducting experiments to examine and characterize this
process.

\end{document}